\documentstyle[psfig,editedvolume]{crckapb}

\def\0134{RX\,J0134-42}
\def\G{$\Gamma_{\rm x}$ }
\def\ros{{\sl ROSAT }}

\def\approxlt{\mathrel{\hbox{\rlap{\lower.55ex \hbox {$\sim$}}
        \kern-.3em \raise.4ex \hbox{$<$}}}}
\def\approxgt{\mathrel{\hbox{\rlap{\lower.55ex \hbox {$\sim$}}
        \kern-.3em \raise.4ex \hbox{$>$}}}}

\begin{opening}
\title{The exceptional X-ray properties of 
        the NLSy1 galaxy \mbox{RX\,J0134.3-4258}} 

\author{Stefanie Komossa}
\author{Dieter Breitschwerdt}
\institute{Max-Planck-Inst. extraterrestrische Physik, Postfach 1603,\\
           85740 Garching, Germany;~~ skomossa@xray.mpe.mpg.de}
\author{Jochen Greiner}
\institute{Astrophys. Institut Potsdam, 14482 Potsdam, Germany}
\author{Janek Meerschweinchen}
\institute{Weststrasse 19, 3063 Obernkirchen 2, Germany}

\end{opening}

\runningtitle{The exceptional X-ray properties of RX\,J0134.3-4258}

\begin{document}


\vspace*{-11cm}
\begin{verbatim}
Contribution to the proceedings of `Astrophysical Dynamics'
(Evora, April 14-16, 1999); to appear in Ap&SS.
Preprint available at http://www.xray.mpe.mpg.de/~skomossa
\end{verbatim}
\vspace*{9.1cm}

\begin{abstract}
The 
Narrow-line Seyfert\,1 galaxy RX\,J0134.3-4258 was detected during
the \ros all-sky survey with an ultrasoft spectrum (Greiner 1996).
Our later pointed observation
led to the discovery of drastic spectral variability.
Here, we present the first detailed analysis of the soft X-ray properties
of this peculiar source.
\end{abstract}

\section{Introduction and data analysis}
\0134 is a Narrow-line Seyfert\,1 (NLSy1) galaxy at redshift $z$=0.237 (Grupe 1996).
It exhibited an ultrasoft spectrum during the \ros all-sky survey (RASS)
in Dec. 1990. Interestingly, the spectrum had changed to flat in
our subsequent pointed observation (Dec. 1992).
The strong spectral variability which
was also noted by Grupe (1996)  
is explored in detail below (for the full analysis, see
Komossa 1999){\footnote{
First results of this study
were presented by Komossa \& Fink (e.g., 1997b)
and Komossa \& Greiner (1999); another study by Grupe et al. (1999)
is underway.}.
Such spectral variability is rare among AGN, and provides
important information on the intrinsic emission mechanism
and/or the properties of surrounding reprocessing material.

\noindent {\em \underline {RASS observation}.}
When fit by a single powerlaw, the spectrum of \0134 turns out to be one
of the steepest among NLSy1s with photon index \G $\approx$ --4.4 (cold absorption was fixed
to the Galactic value,  
$N_{\rm Gal} = 1.59\,10^{20}$ cm$^{-2}$).
One efficient mechanism to produce such a steep X-ray
spectrum 
is the presence of a  
`warm absorber', highly ionized material found in the nuclei of
about 50\% of the Seyfert galaxies (e.g., Komossa \& Fink 1997a,b). 
There are also indications that the flat-state spectrum is still
mildly warm-absorbed (Fig. 1). 
In fact, a warm-absorbed, intrinsically {\em flat} powerlaw provides a successful
alternative fit to the RASS data.
A large
column density $N_{\rm w}$ (of the order 10$^{23}$ cm$^{-2}$) is needed to
account for the ultrasoft observed spectrum.
When we fix \G = $-2.2$, the value observed during the later pointing,
and use  $N_{\rm H}$ = $N_{\rm Gal}$, we obtain $\log N_{\rm w} = 23.1$ and 
an ionization parameter $\log U = 0.5$.
This model gives an excellent fit ($\chi{^{2}}_{\rm red}=0.6$).
A number of further models were applied, some
summarized in Tab. 1.

\begin{table*} 
  \caption{Comparison of different spectral fits to \0134:
(i) single powerlaw (pl),
(ii) accretion disk model after Shakura \& Sunyaev, and (iii) warm absorber.
  \G was fixed to --2.2 in (ii) and (iii). `S' = RASS observation, `P' = pointing. 
 }
  \begin{tabular}{llllllllll}
  \noalign{\smallskip}
  \hline
  \noalign{\smallskip}
   obs. &  \multicolumn{3}{l}{powerlaw$^{(1)}$~~~~~~~~~~~~} &
                       \multicolumn{3}{l}{acc. disk + pl$^{(1)}$~~~~~~} &
                         \multicolumn{3}{l}{warm absorber$^{(1)}$} \\
  \noalign{\smallskip}
         & $N_{\rm Gal}^{(2)}$ & \G~ & $\chi^2_{\rm red}$~~~ & $M_{\rm BH}^{(3)}$ &
                                       ${\dot M}\over{{\dot M_{\rm edd}}}$ &
                                       $\chi^2_{\rm red}$~~~ &
                                       log $U$~ & log $N_{\rm w}$ & $\chi^2_{\rm red}$ \\
  \noalign{\smallskip}
  \hline
    S & 0.16 & --4.4 & 0.5 & ~~1 & 0.1 & 0.6 &~~0.5 & 23.1 & 0.6 \\
    P & 0.16 & --2.2 & 1.4 &  &  &  & &  &  \\
  \hline
  \noalign{\smallskip}
     \end{tabular}
  \label{tabn}

  \noindent{\small
  $^{(1)}$$N_{\rm H}$
fixed to $N_{\rm Gal}$ ~ $^{(2)}$in 10$^{21}$cm$^{-2}$ ~ $^{(3)}$ black 
hole mass in 10$^5$M$_{\odot}$, fixed~
 }
\end{table*}

\noindent {\em \underline {Pointed observation}.}
The fit of a single powerlaw to the spectrum of \0134
yields a photon index \G = $-2.2$ ($\chi^2_{\rm red}$ = 1.4),
much flatter than during the RASS observation.
For this model fit, two kinds of residuals are apparent (Fig. 1):
(i) the first data point (below 0.15 keV) indicates a higher
countrate than predicted by the model. This data point
significantly influences the value of $\chi^2_{\rm red}$, and
if it is excluded from spectral fitting, we obtain $\chi^2_{\rm red}$ = 1.0
and \G = $-2.1$.
Formally, a very low-temperature soft excess could be present
in the spectrum of \0134 but was not fit since constrained by only
one data point. 
The second deviation from the powerlaw is (ii) an underprediction
of the countrate in the energy range $\sim$0.4--0.9 keV (Fig. 1)
indicative of the presence of absorption edges,
as observed in AGNs where warm absorbers are present.
However, again, the deviations from the powerlaw are only defined by
few bins, and we thus assume in the following
that the spectrum during the pointed observation essentially
represents the intrinsic, un-distorted continuum.

\noindent {\em \underline {Temporal analysis}:}
The mean countrate is nearly constant from RASS (0.30 cts/s) to
pointed (0.24 cts/s) observation. During the latter, the lightcurve
(see Fig. 2 of Komossa \& Greiner 1999) reveals variability by a factor $\sim$2.

\begin{figure*}[ht]
     \vbox{\psfig{figure=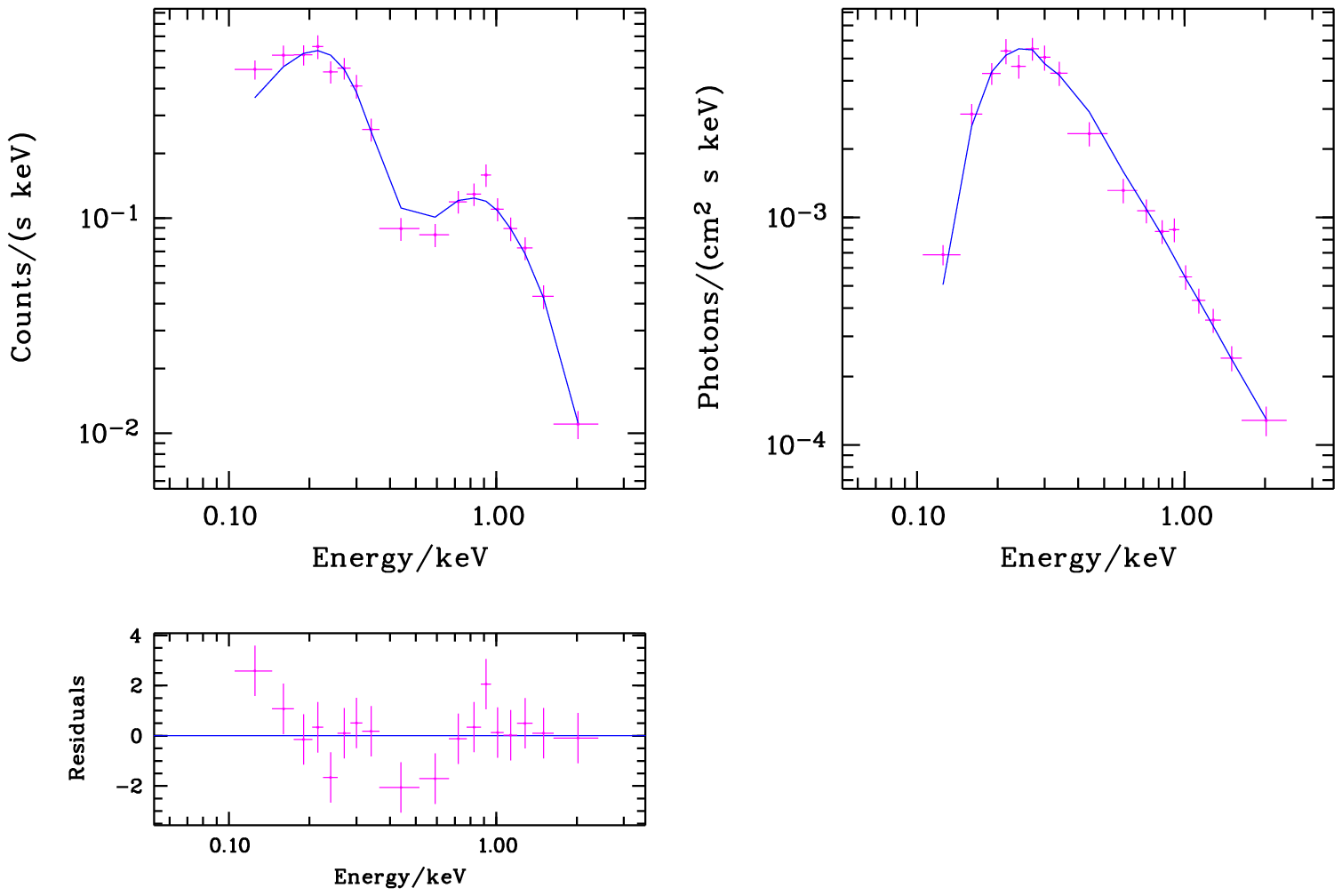,width=4.9cm,
          bbllx=2.1cm,bblly=5.8cm,bburx=10.1cm,bbury=11.7cm,clip=}}\par
    \vbox{\psfig{figure=komossa3_fig1.ps,width=4.9cm,%
     bbllx=2.1cm,bblly=1.1cm,bburx=10.1cm,bbury=4.4cm,clip=}}\par    
    \vspace*{-5.4cm}\hspace*{4.8cm}
     \vbox{\psfig{figure=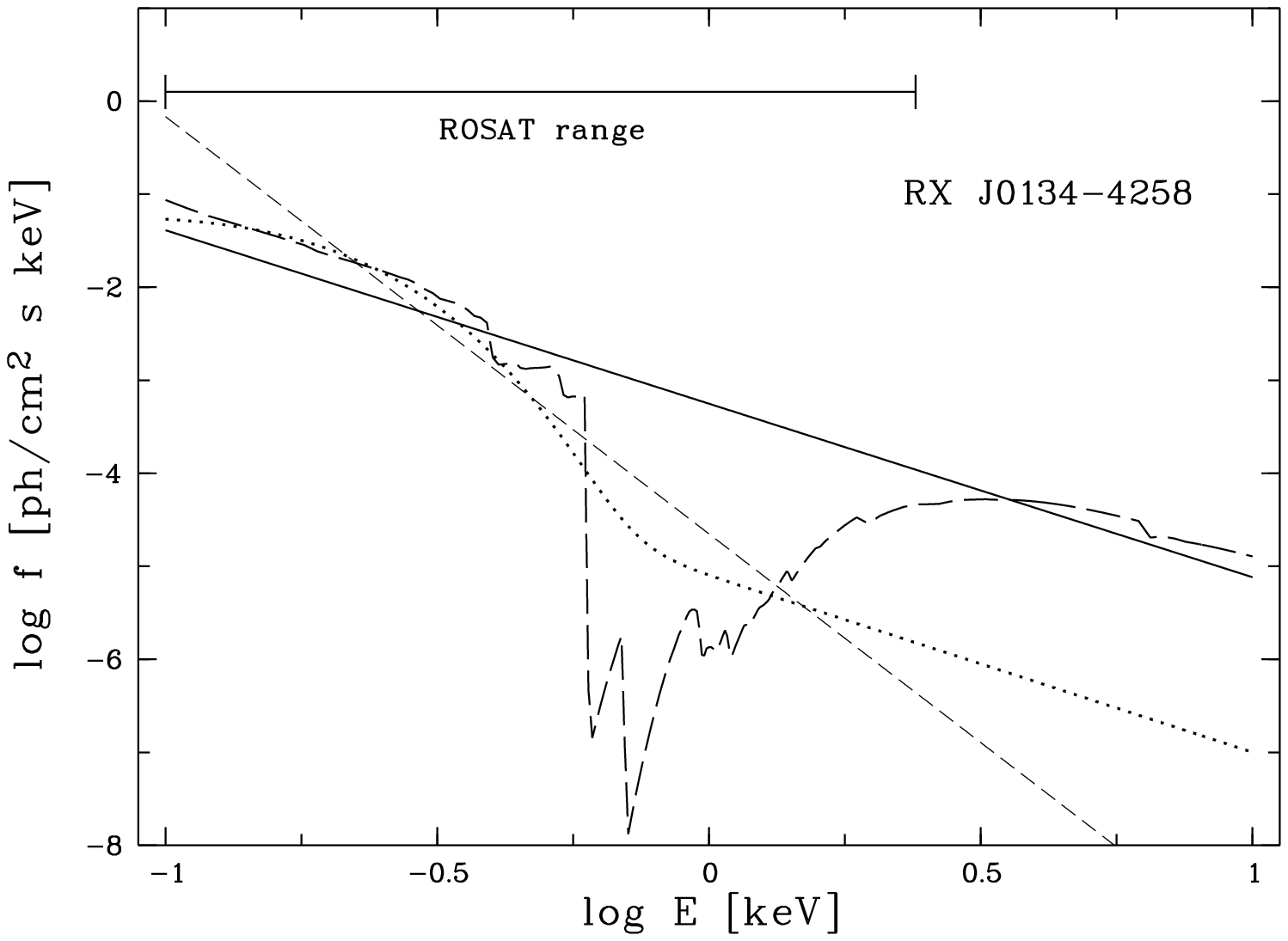,width=7.5cm,%
          bbllx=1.5cm,bblly=1.1cm,bburx=18.3cm,bbury=12.2cm,clip=}}\par
  \caption[]{{\em Left}:  
The upper panel gives the observed X-ray spectrum (crosses) of \0134 
(pointed observation) and powerlaw model
fit (solid line), the lower panel displays the residuals.
{\em Right}: 
Comparison of different spectral models (corrected for detector response)
sucessfully fit to the \ros survey
observation of \0134 (broken lines) and the pointed observation (solid line).
Short-dashed line: single powerlaw with $\Gamma_{\rm x}=-4.4$; long-dashed: warm-absorbed
flat powerlaw; dotted: powerlaw plus soft excess, parameterized as a black body.
}
  \label{fits_0134}
\end{figure*}

\section{Discussion: Variability mechanisms}

\paragraph{Warm absorption.}

One mechanism that might explain the spectral variability in \0134
is warm absorption because this is an efficient way to
produce variable, and steep X-ray spectra (e.g., Komossa \& Fink 1997a,b).
Note, that Grupe (1996) argued against
the presence of a warm absorber based on the erroneous statement
that warm absorption could not produce a steep soft X-ray spectrum.
Examination whether, and under which conditions, a warm absorber is
indeed a viable description of the
X-ray spectrum, and whether it is the only one,
has to be based upon detailed modeling and careful consideration of
alternatives:  

The most suggestive scenario within the framework of warm absorbers
is a change in the {\em ionization state} of matter along the line of sight,
caused by {\em varying irradiation} by a central ionizing source.
One problem arises immediately, though:
In the simplest case, {\sl lower} intrinsic luminosity would 
be expected to cause the {\sl deeper}  
observed absorption in 1990. However,
the source is somewhat {\sl brighter} in the RASS observation.
If one wishes to keep this
scenario, one would have to assume that the ionization state of the absorber still reflects
a preceding (unobserved) low-state in intrinsic flux.

Alternatively, gas heated by the central continuum source
may have {\em crossed the line of sight},
producing the steep RASS spectrum, and has (nearly) disappeared in the 1992 observation.
This scenario explains most naturally the nearly constant countrate
from RASS to pointed observation, because the countrate is dominated
by the soft energy part of the spectrum (below 0.7 keV) which is
essentially unaffected by warm absorption.
The transient passage of a BLR cloudlet would be consistent with the
scenario proposed by Rodriguez-Pascual et al. (1997) who suggested 
the presence of {\em matter-bounded}
BLRs in NLSy1 galaxies to account for some of their peculiar optical
properties.

\noindent{\em Alternatives:~~}
(I.) It is important to keep in mind the short duration of the RASS observation,
and both, an intrinsically steep powerlaw and a strong soft excess fit the
X-ray spectrum as well.
Variability in only one component seems to be problematic, though, since the
nearly constant countrate has to be accounted for.
(II.) A spectral change with constant countrate is reminiscent of one class
of Galactic black hole (BH) transients.  
In fact, the potential similarity of NLSy1s with Galactic BH 
candidates has been repeatedly mentioned 
but has never been explored in more detail (see Komossa 1999 for
more comments on this possibility). 
(III.) Finally, it is also possible that the constant countrate is pure
coincidence: Both, variable soft excesses 
{\em and} variable powerlaws (often of constant shape)
have been observed in AGN and these two might have compensated
each other to produce nearly constant total countrate. 

\section {Summary and conclusions}

\0134 underwent a drastic X-ray spectral
transition from steep
(\G $\simeq -4.4$) to flat (\G $\simeq -2.2$) between
our two \ros observations separated by 2 yr, while
the mean countrate remained nearly constant. We
examined several scenarios that might account for
this peculiar behavior, with focus on the presence of a warm
absorber. We find that a reaction
of the ionized material to continuum changes requires
non-equilibrium effects to be at work.
Alternatively, and more likely,
a cloud of warm gas may have passed our line of sight.
Variability of both components in the framework of a
powerlaw-plus-soft-excess spectral description provides an alternative explanation.
High spectral resolution observations at soft X-ray energies will provide
further clues on the nature of this interesting source.

{}

\end{document}